\documentclass[12pt]{iopart}

\expandafter\let\csname equation*\endcsname\undefined
\expandafter\let\csname endequation*\endcsname\undefined
\usepackage{amsmath}
\usepackage{color}
\usepackage{subcaption}

\usepackage{microtype}
\usepackage{graphicx}
\usepackage{booktabs} 

\usepackage{amssymb}
\usepackage{mathtools}
\usepackage{amsthm}

\usepackage{hyperref}
\usepackage{here}

\begin{document}

\title[NN Soliton]{Comparative Study of Neural Network Methods for Solving Topological Solitons
}

\author{Koji Hashimoto$^1$, Koshiro Matsuo$^2$,
Masaki Murata$^2$, Gakuto Ogiwara$^2$}
\address{$^1$ Department of Physics, Kyoto University, Kyoto 606-8502, Japan}
\address{$^2$ Department of Information Systems,  Saitama Institute of Technology, Saitama 369-0293, Japan}
\ead{koji@scphys.kyoto-u.ac.jp, f3011hxs@sit.ac.jp, m.murata@sit.ac.jp, f3002keg@sit.ac.jp}
\vspace{10pt}

\begin{abstract}
Topological solitons, which are stable, localized solutions of nonlinear differential equations, are crucial in various fields of physics and mathematics, including particle physics and cosmology. However, solving these solitons presents significant challenges due to the complexity of the underlying equations and the computational resources required for accurate solutions. To address this, we have developed a novel method using neural network (NN) to efficiently solve solitons. A similar NN approach is Physics-Informed Neural Networks (PINN). In a comparative analysis between our method and PINN, we find that our method achieves shorter computation times while maintaining the same level of accuracy. This advancement in computational efficiency not only overcomes current limitations but also opens new avenues for studying topological solitons and their dynamical behavior.
\end{abstract}

%
%
%
%
%

\section{Introduction}

Topological solitons are fascinating mathematical constructs with profound implications across various domains of physics and mathematics. These stable, localized solutions of nonlinear differential equations have been pivotal in advancing our understanding of complex phenomena ranging from particle physics to cosmology \cite{MantonSutcliffe2004}. The presence of solitons in these fields often provides comprehensive understanding of the dynamics of the system, enabling us to model and interpret physical events that perturbative approaches may fail to elucidate \cite{DrazinJohnson1989}.

However, the quest to accurately solve the equations governing topological solitons poses significant challenges. The challenge in solving solitons is that analytical solutions are often not available due to nonlinearity and complexity of the equations, so numerical methods has to be relied upon \cite{Scott1999}. In addition, conventional numerical methods consume large amounts of computational resources and require enormous computation time to perform highly accurate analysis. 

These challenges underscore the need for efficient methods capable of handling the inherent complexity without compromising on precision. For more than two decades, it has been expected that differential equations can be solved efficiently using NN, supported by the universal approximation theorem \cite{HORNIK1989359}. This capability has further advanced with the development of novel optimization methods in machine learning \cite{chen2018neural}. Such advancements have led to successful applications in solving physical systems, including the Schrödinger equation in quantum mechanics \cite{saito2018method,pfau2020ab,hermann2020deep,naito2023multi} and even the holographic principle for gravity \cite{hashimoto2018deep}. 

In recent years, the advent of NN has revolutionized various scientific domains, offering novel approaches to age-old problems.  
PINN have emerged as one such powerful tool by embedding physical laws directly into the NN architecture \cite{RaissiPerdikarisKarniadakis2019}. PINN is particularly effective in describing physical phenomena because it can incorporate physical laws into the learning process when finding the solution of differential equations. 
Research on PINN has explored various strategies to improve computational speed \cite{bafghi2023pinnstf2fastuserfriendlyphysicsinformed}.
However, it remains unclear how effective NNs, including PINNs, are in finding soliton solutions.
Therefore, we developed our own NN model and conducted a comparative analysis with PINN, focusing on computation time and accuracy.
Our analysis demonstrates that our model significantly reduces computation time while maintaining accuracy comparable to that of PINN.
This advancement not only mitigates current computational constraints but also paves the way for more extensive and dynamic analyses of topological solitons.

In the following sections, we first detail the construction and implementation of our NN method tailored for soliton solutions. 
Next, we provide an in-depth comparison of our approach with PINN, highlighting improvements in computational efficiency and accuracy. 
Finally, we discuss the broader implications of our findings and potential future applications, emphasizing the transformative impact on the study of topological solitons.



\section{Method}

In this section, we introduce our NN method, which we call Neural Network for Difference Equation (NNDE). 
The objective of this model is to solve a differential equation with the following form:
\begin{align}
D(x,f(x),f'(x),\cdots) = 0,
\label{eq:DF}
\end{align}
where $\cdots$ in $D$ represents higher-order derivatives of $f(x)$. The function $f(x)$ is sufficiently differentiable with respect to the coordinate $x$ to meet the conditions of the differential equation.
While we focus on the problem to find a single function $f(x)$ with respect to one dimensional space coordinate $x$, the application to higher dimensions and multiple functions is straightforward.
Additionally, $f(x)$ must satisfy the boundary conditions of the form:
\begin{align}
B_b(x_b, f(x_{b}), f'(x_b), \cdots) = 0 ,
\label{eq:BC}
\end{align}
where $b$ is the index of the boundary conditions, $x_b \ (b=1,2,\cdots)$ are the points at which the boundary conditions are imposed, and $\cdots$ in $B_b$ similarly includes higher-order derivatives. 
Note that our method is also applicable to non-local boundary conditions, such as periodic boundary conditions.

\subsection{NNDE}

\begin{figure}[t]
   \begin{center}
      \includegraphics[scale=0.7]{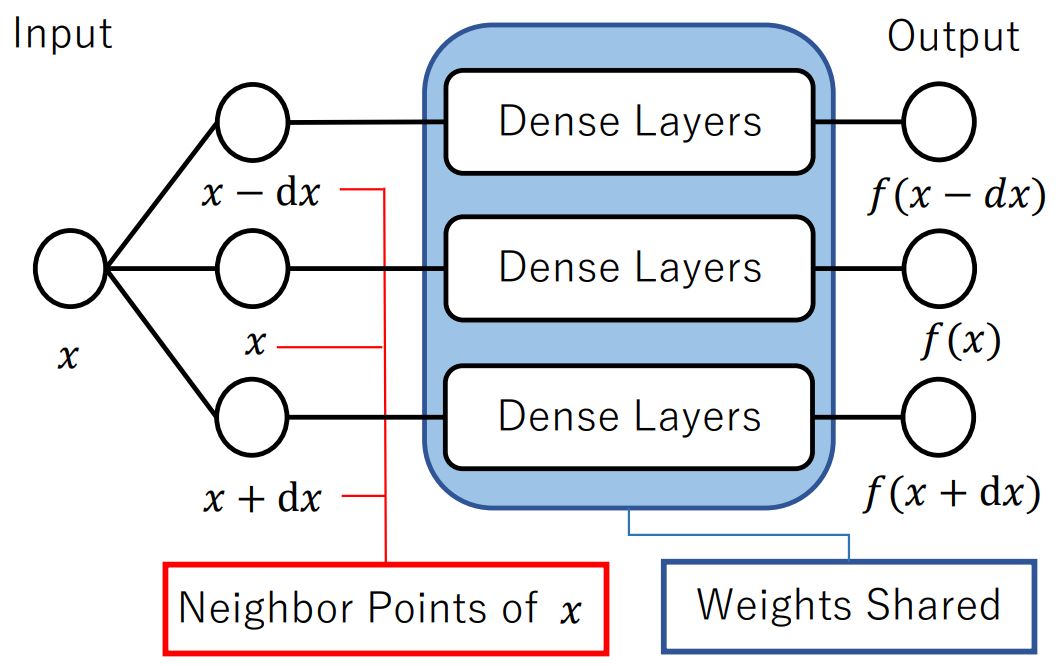}
      \caption{Structure of NNDE.}
      \label{fig:NNstructure}
   \end{center}
\end{figure}

We illustrate NNDE model for solving the differential equation \eqref{eq:DF} with the boundary conditions \eqref{eq:BC}.
The structure of the NNDE designed to solve second-order differential equations is illustrated in Figure \ref{fig:NNstructure} as an example. 
This framework can be extended to accommodate higher-order cases as well.
The details of each layer in NNDE are as follows: 
\begin{enumerate}
  \item The first layer is the input layer, which has one unit that receives $x$ as input.
  \item The second layer receives $x$ from the first layer and outputs a vector whose components are $x$ and the neighborhood points of $x$.
  For example, $(x+dx, x, x-dx)$ is output when solving a second-order differential equation, where $dx$ is treated as a hyperparameter. 
 \item The following layers are made up of dense layers. Each of $x$ and its neighboring points, as output from the second layer, is passed through separate dense layers. All dense layers share the same weights. 
  \item In the output layer, the solution functions corresponding to $x$ and its neighboring points are produced from each dense layer. 
\end{enumerate}
In practice, the dense layers in the figure comprise four layers with 32, 64, 32, and 1 units, respectively, using the $\tanh$ function as the activation function. 
The summary of this model is shown in Table \ref{model-table}.
The right column of the table displays the output shapes for each layer, where "NONE" indicates the batch size.
\begin{table}[t]
\caption{Summary of NN model.}
\label{model-table}
\vskip 0.15in
\begin{center}
\begin{small}
\begin{sc}
\begin{tabular}{ccc}
\toprule

Layers & Output Shape  \\
\midrule
First Layer & (None, 1, 1)   \\
Second Layer & (None, 3, 1)  \\
Dense & (None, 3, 32)  \\
Dense & (None, 3, 64)  \\
Dense & (None, 3, 32)  \\
Dense & (None, 3, 1)  \\
\bottomrule
\end{tabular}
\end{sc}
\end{small}
\end{center}
\vskip -0.1in
\end{table}

The loss function is defined as the sum of the square of the differential equation and the squares of the boundary conditions:
\begin{align}
    L &=  L_{\rm DF} + L_{\rm BC}\nonumber
    \\
    L_{\rm DF} &= D(x, f(x), f'(x), \cdots)^2\nonumber ,
    \\
    L_{\rm BC} &= 
    \sum_{b} B_b(x_b,f(x_b),f'(x_b),\cdots )^2 .
    \label{eq:loss}
\end{align}
In $L_{\rm DF}$ and $L_{\rm BC}$, the derivatives $f'(x), f''(x),\cdots$ are discretized:
\begin{align}
   f'(x) &\equiv \frac{f(x+dx)-f(x-dx)}{2dx} ,\\
   f''(x) &\equiv 
   \frac{f(x+dx)+f(x-dx) - 2f(x)}{dx^2} .
\end{align}

A key distinction between NNDE and PINN lies in the method used for calculating derivatives. NNDE employs a difference method for approximating derivatives, where finite differences are used to estimate the derivatives of the solution function. 
In contrast, PINN utilizes automatic differentiation (autodiff) to compute derivatives directly within the NN framework. 
Autodiff leverages the backpropagation and is capable of calculating exact gradients with respect to network parameters. 
This method is often more flexible in handling continuous functions, as it does not rely on spatial discretization. 
However, since autodiff involves backpropagation computation not only for the gradient of the loss function but also for the derivatives of the function, PINN may result in longer computation time.
In the following sections, we shall compare the accuracy and the computation time of NNDE with those of PINN for some nonlinear field theories that possess soliton solutions. 

The NN models are implemented in this study with the use of  TensorFlow. 
The following hyperparameters are commonly used in this paper: optimizer $=$ Adam, $dx = 10^{-2}$, epochs $= 10000$.

\subsection{\label{sec:level2}Field equations}
In this section, we describe the field equations to which we will apply NN models: the $\phi^4$ theory and the Sine-Gordon model.
Both field theories are well-known for possessing a specific type of soliton solution call the kink soliton.

\subsubsection{$\phi^4$ theory}

The differential equation for the $\phi^4$ theory in one-dimensional space is given by
\begin{align}
   f''(x) + m^2 f(x) - \lambda f(x)^3 = 0 .
\end{align}
Here, $m^2$ and $\lambda$ are mass sqauared and the coupling constant respectively, and we assume time-independence.
In this theory, there are two distinct stationary points in the field space:
$f_{\pm} = \pm \sqrt{m^2/\lambda}$.
It is well-known that this theory admits a topological kink solution that interpolates between these stable states, expressed as 
\begin{align}
   f_{\rm kink}(x) = f_+ \tanh \left[\frac{m}{\sqrt{2}}(x-a)\right] \,,
\end{align}
where $x = a$ is the position of the kink center. 
In fact, this solution asymptotically approaches the stationary points such that $f(x=\pm\infty)=f_\pm$.
For numerical calculations, it is inefficient to consider the entire real axis of $x$, so we apply a coordinate transformation $\tilde{x}=\tanh(x)$, mapping $\mathbb{R}$ to $[-1,1]$. 
The equation of motion in terms of $\tilde{x}$ is 
\begin{align}
   (1-\tilde{x}^2)^2\frac{d^2 f }{d\tilde{x}^2}
   - 2\tilde{x}(1-\tilde{x}^2)\frac{df }{d\tilde{x}}
   + m^2 f - \lambda f^3
   = 0 ,
\end{align}
In addition, the asymptotic behavior $f(\pm \infty)=f_{\pm}$ is reformulated as the boundary conditions:
\begin{align}
   f(\pm 1) = f_\pm\,, 
   \quad
   f(0)=0 .
\end{align}
Here, the condition $f(0)=0$ is imposed to specify the center of mass of the kink. 
For simplicity, in the following we will denote $\tilde{x}$ as $x$. 

To summarize, $D$ and $B_b$ in \eqref{eq:loss}, which contributes to the loss function, are given by
\begin{align}
   D &= (1-x^2)^2 f''(x)
   - 2x(1-x^2) f'(x) 
   \nonumber \\
   & \quad + m^2 f(x) - \lambda f(x)^3 ,
   \\
   B_b &= f(x_b)-f_b ,
\end{align}
where $(x_b,f_b) = (-1,f_-),(0,0),(1,f_+)$. 
In our numerical computation, we set $m=\lambda=1$.

\subsubsection{Sine-Gordon equation}

The time-independent form of the Sine-Gordon equation is given by 
\begin{align}
   f''(x) - \sin f(x) = 0 .
\end{align}
There are series of stationary points expressed as $f(x)=2\pi n$.
Similar to the $\phi^4$ case, this equation admits a topological kink solution of the form 
\begin{align}
   f(x) = 4 \arctan e^{x-a} ,
\end{align}
where $x = a$ represents the position of the kink center. 
This solution interpolates $f=0$ and $f=2\pi$ such that the asymptotic behaviors are $f(-\infty)=0$ and $f(\infty)=2\pi$. 
Again, to perform numerical calculations efficiently, we use the coordinate transformation $\tilde{x} = \tanh(x)$, which maps the entire real line $\mathbb{R}$ to the interval $[-1, 1]$.
Rewriting the equation in terms of $\tilde{x}$ yields: 
\begin{align} 
    (1 - \tilde{x}^2)^2\frac{d^2 f}{d\tilde{x}^2}
    -2\tilde{x}(1 - \tilde{x}^2)\frac{df}{d\tilde{x}}
    -\sin f = 0. 
\end{align}
The boundary conditions for the kink are set as: 
\begin{align} 
f(-1) = 0, \quad f(0) = \pi, \quad f(+1) = 2\pi. 
\end{align} 
Here, $f(0) = \pi$ specifies the center of the kink. 
Again, for simplicity, we replace $\tilde{x}$ with $x$ in the following equations.
The contributions to the loss function for this problem are defined as: 
\begin{align}
   D &= (1-x^2)^2 f''(x)
   - 2x(1-x^2) f'(x) 
   \nonumber 
   - \sin f(x) ,
   \\
   B_b &= f(x_b)-f_b ,
\end{align}
where the boundary points are $(x_b, f_b) = (-1, 0), (0, \pi), (1, 2\pi)$.

\subsection{Training condition}
We use both NNDE and PINN  methodologies for the $\phi^4$ theory and the Sine-Gordon equation, both of which have known exact analytical solutions. 
We compare the NNDE results with these exact solutions to assess the accuracy of our approach. 
Additionally, we perform a similar evaluation with PINN to determine the accuracy of its results. 
Lastly, we compare the computational efficiency of NNDE and PINN in terms of both computational time and accuracy.

As depicted in Figure \ref{fig:NNstructure}, our model uses unsupervised learning such that the training data consists only of the input $x$, which is the coordinate value.
The training and test data use values ranging from $-1.0$ to $1.0$ for both the $\phi^4$ theory and the Sine-Gordon equation. 
For the test data, we use values ranging from $-1.0$ to $1.0$ with a discretization step size of $10^{-2}$.
On the other hand, we use various discretization step sizes for the training data and various batch sizes to ensure a comprehensive comparison between NNDE and PINN.
The discretization step sizes for the training data are set to $10^{-1}$, $10^{-2}$, $10^{-3}$, $10^{-4}$, and $10^{-5}$, and the batch sizes are set to $10\%, 20\%, 25\%,$ and $50\%$ of the total data points. 

For each combination of step size and batch size, we perform the training and measure two metrics: the Mean Squared Error (MSE) between the model's predictions and the exact solutions for the test data, and the computation time for training. 
We then average the MSE and computation time results across all combinations to provide a comprehensive comparison.
We compare the predictions of each model with the exact solutions for this test data to evaluate their accuracy.

The computations were performed on a machine equipped with an NVIDIA GeForce RTX 3060 GPU, 64.0GB of memory, and a Intel Core i7-12700 CPU. 
The environment was constructed using Docker, with TensorFlow 2.9.2 as the primary library.

\section{Results}
Here we present a comprehensive comparison between NNDE and PINN for both $\phi^4$ theory and Sine-Gordon equation.


\subsection{Kink soliton of $\phi^4$ theory}
The exact kink solution for $m=\lambda=1$ to be compared to the NN predictions is
\begin{align}
   f_{\rm true}(x) = \tanh \left[\frac{1}{\sqrt{2}}(\tanh^{-1} x)\right] .
\end{align}
We compare each model's predictions with this exact solution to verify their accuracy.

\begin{figure}[t]
    \centering
    \begin{subfigure}{0.48\linewidth}
        \centering
        \includegraphics[scale=0.5]{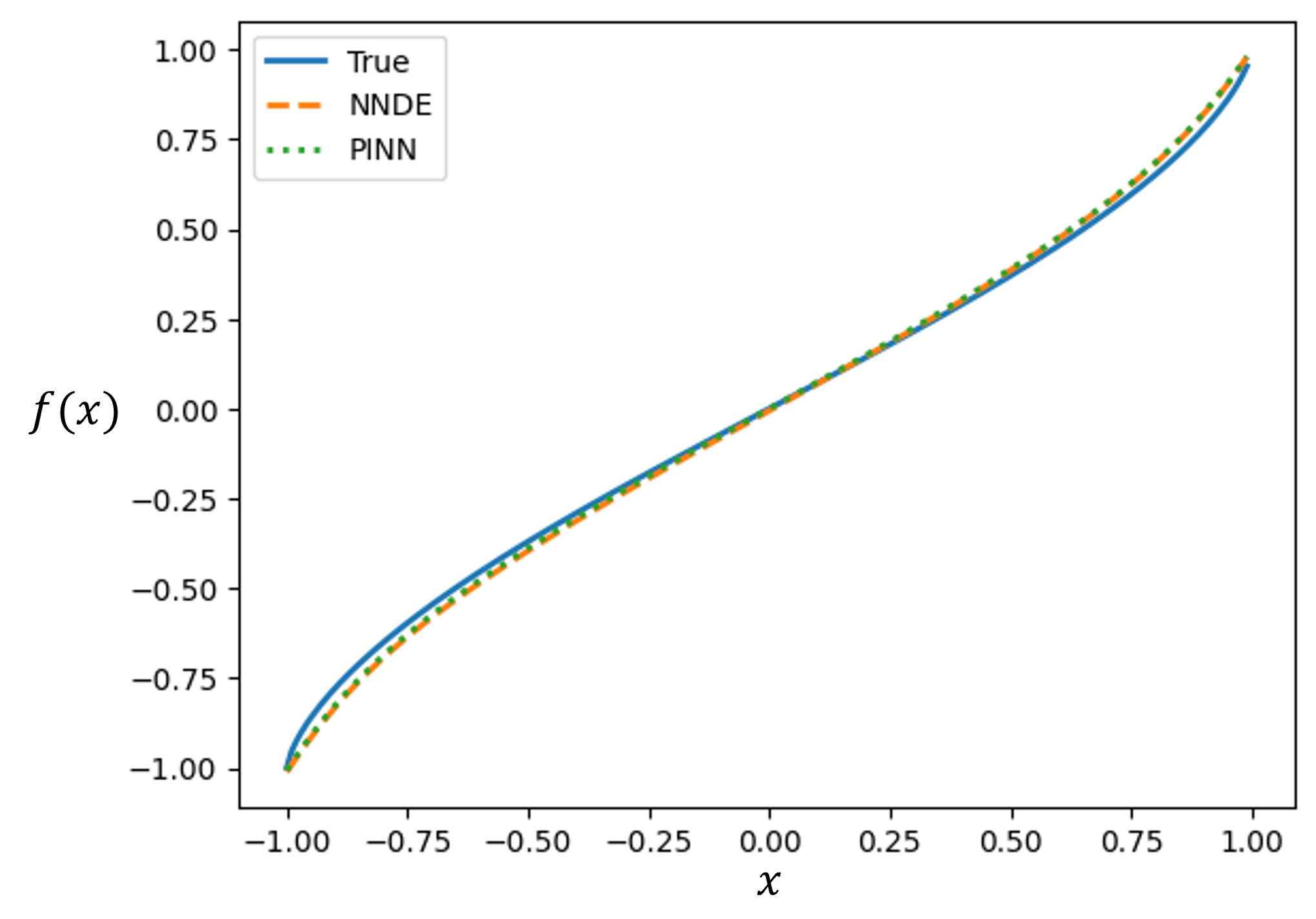}
        \caption{$\phi^4$ theory}
        \label{fig:phi4_pred}
    \end{subfigure}
    \hfill 
    \begin{subfigure}{0.48\linewidth}
        \centering
        \includegraphics[scale=0.5]{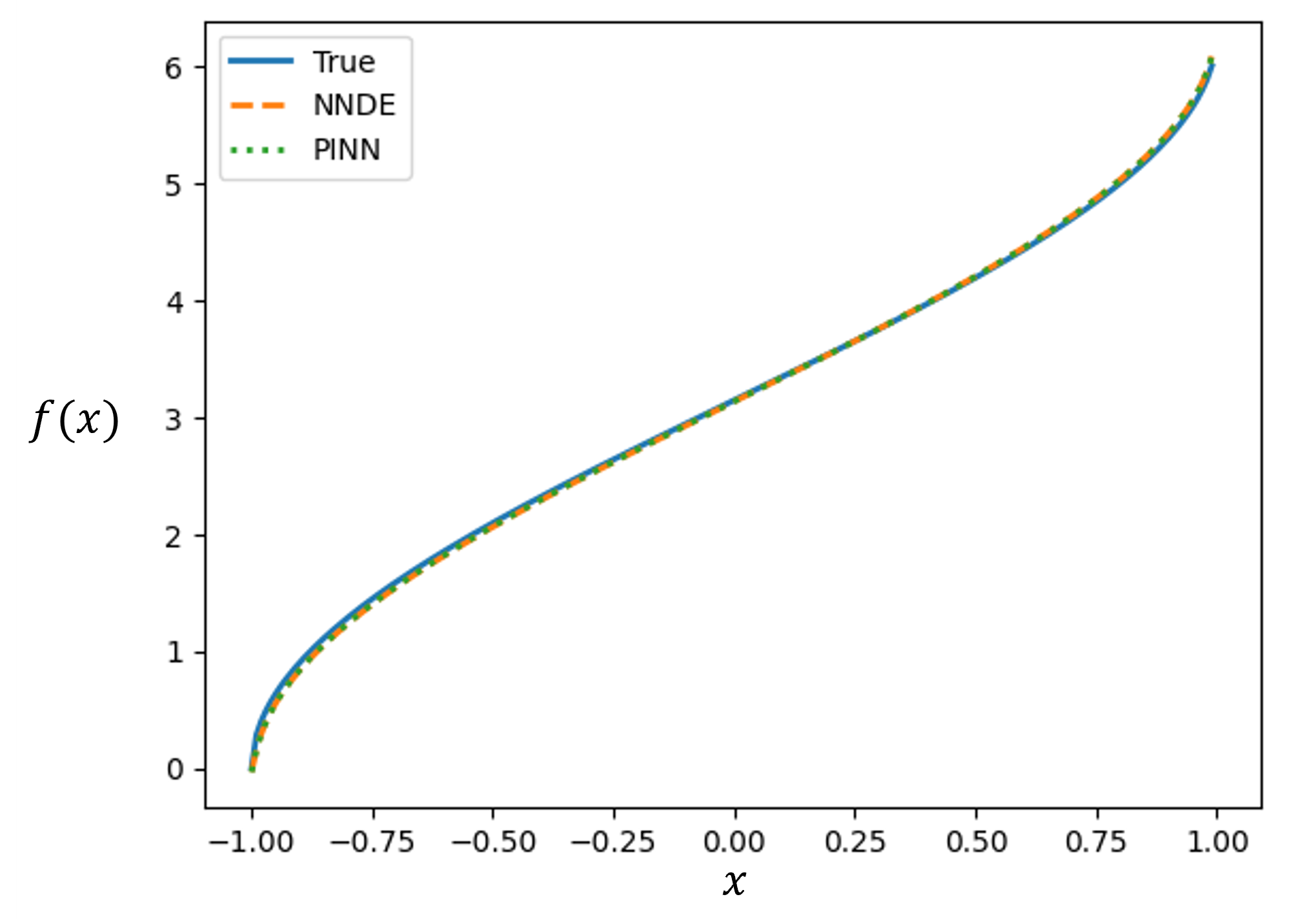}
        \caption{Sine-Gordon equation}
        \label{fig:SG_pred}
    \end{subfigure}
    \caption{Comparison of the exact solution, NNDE, and PINN predictions with step size $10^{-5}$ and batch size 10\%.}
\end{figure}

\begin{table}[H]
\caption{Comparison of average accuracy (MSE) and computation time for the $\phi^4$ theory and the Sine-Gordon equation.}
\label{compare-table}
\vskip 0.15in
\begin{center}
\begin{small}
\begin{sc}
\begin{tabular}{c|cc|cc}
\toprule
 & \multicolumn{2}{c|}{$\phi^4$} & \multicolumn{2}{c}{Sine-Gordon} \\
\midrule
 & MSE & Time & MSE & Time \\
\midrule
NNDE & 0.002333 & 1786.555 & 0.014344 & 1702.768 \\
PINN & 0.002192 & 3293.328 & 0.015208 & 3144.233 \\
Difference (absolute value) & 0.000141 & 1506.773 & 0.000864 & 1441.465 \\
Difference (relative value) & 0.064339 & 0.457523 & 0.056808 & 0.458447 \\
\bottomrule
\end{tabular}
\end{sc}
\end{small}
\end{center}
\vskip -0.1in
\end{table}

\begin{figure}[t]
    \centering
    \begin{center}
    \begin{subfigure}{0.88\linewidth}
        \centering
        \includegraphics[scale=0.55]{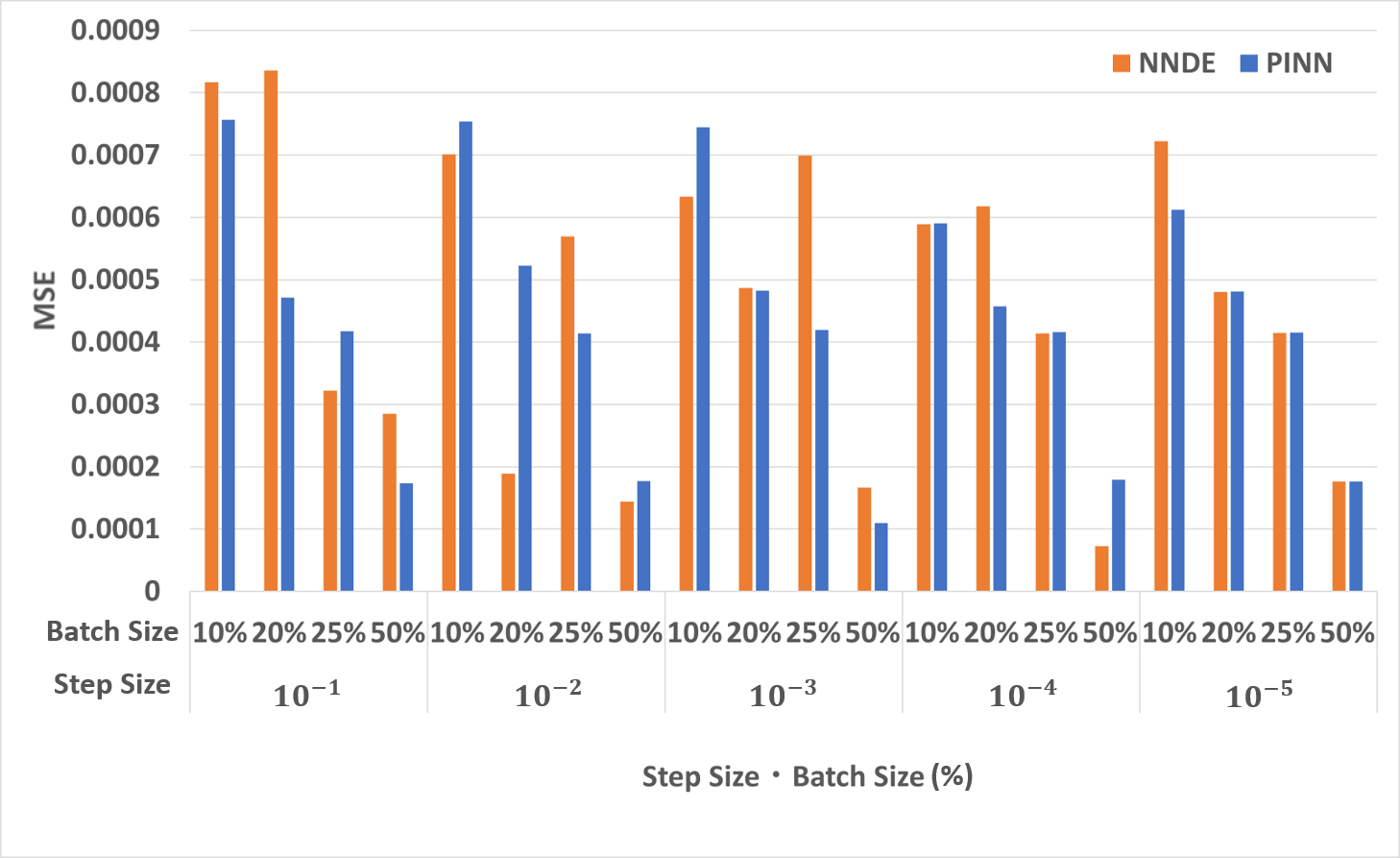}
        \caption{MSE}
        \label{fig:phi4_mse}
    \end{subfigure}
    \end{center}
\vspace{5mm}
\centering
    \begin{subfigure}{0.88\linewidth}
        \centering
        \includegraphics[scale=0.55]{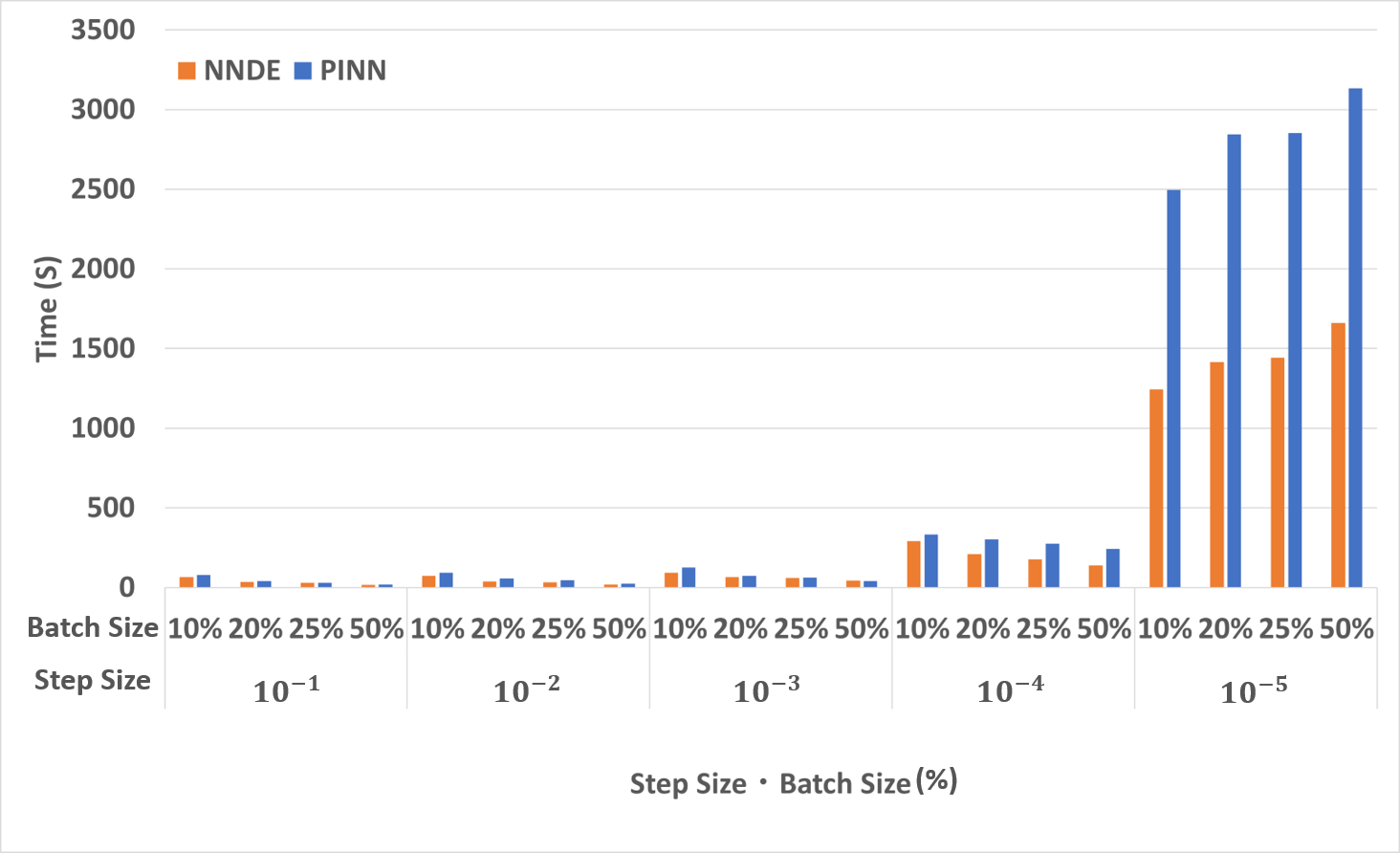}
        \caption{Computation time}
        \label{fig:phi4_time}
    \end{subfigure}
    \caption{Comparison of MSE and computation time for the $\phi^4$ theory with varying step sizes and batch sizes}
    \label{fig:phi4_combined}
\end{figure}

Figure \ref{fig:phi4_pred} shows a comparison between the exact solution and the predictions of NNDE and PINN, where the step size is $10^{-5}$ and the batch size is $10\%$. 
This figure shows that both of NNDE and PINN provide a reasonable approximation.
Similar results are observed for different choices of step size and batch size.
As summarized in Table \ref{compare-table}, NNDE achieved an average MSE of $0.002333$, which was comparable to the MSE of $0.002192$ obtained with PINN, with a relative difference of only $6.43\%$. 
However, the computation time for NNDE was significantly lower, averaging 1786.555 seconds, compared to $3293.328$ seconds for PINN, representing an {45.75\%} reduction in computation time. 
This trend was consistent across various step sizes and batch sizes, as illustrated in Figure \ref{fig:phi4_combined}, where subfigure \ref{fig:phi4_mse} shows the MSE and subfigure \ref{fig:phi4_time} displays the computation time for different configurations. 
These results demonstrate that NNDE is more time-efficient than PINN for $\phi^4$ theory.


\subsection{Kink soliton of Sine-Gordon equation}
\begin{figure}[t]
    \centering
    \begin{center}
    \begin{subfigure}{0.88\linewidth}
        \centering
        \includegraphics[scale=0.55]{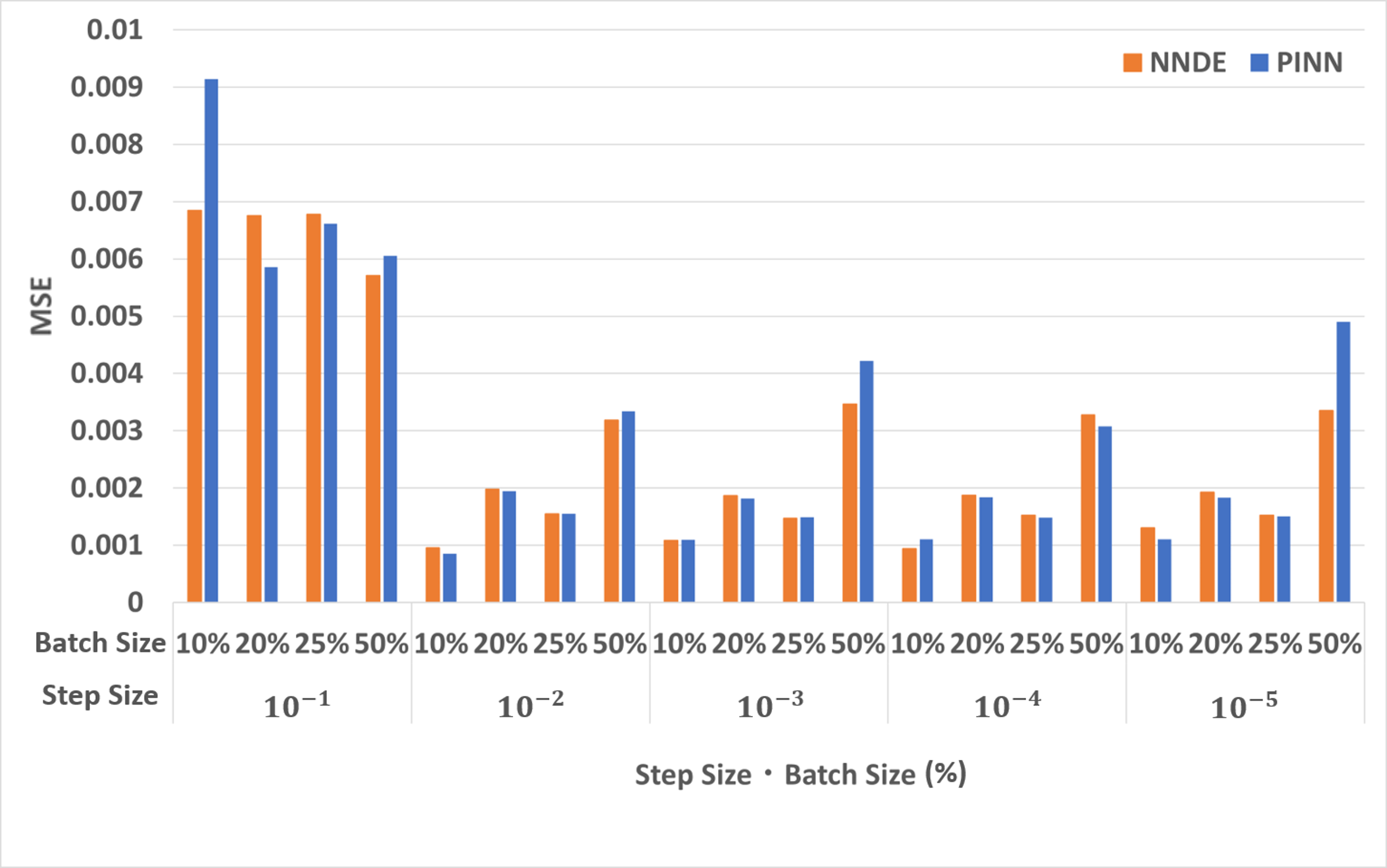}
        \caption{MSE}
        \label{fig:SG_mse}
    \end{subfigure}
    \end{center}
\vspace{6mm}
    \begin{subfigure}{0.88\linewidth}
        \centering
        \includegraphics[scale=0.55]{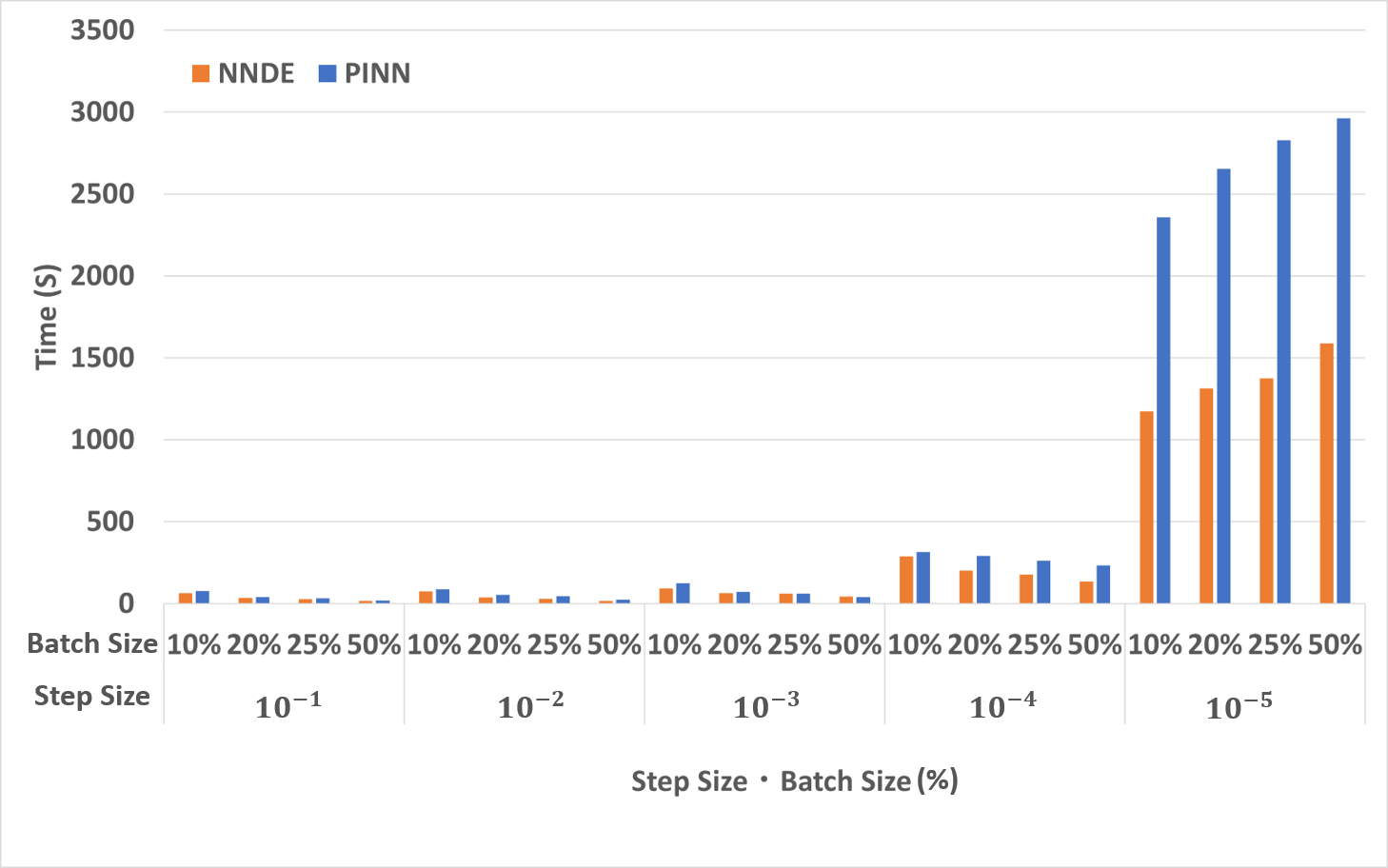}
        \caption{Computation time}
        \label{fig:SG_time}
    \end{subfigure}
    
    \caption{Comparison of MSE and computation time for the Sine-Gordon equation with varying step sizes and batch sizes}
    \label{fig:SG_combined}
\end{figure}

The exact form of the kink solution of the Sine-Gordon equation is 
\begin{align}
   f_{\rm true}(x) = 4 \arctan(\exp(\tanh^{-1}(x))) .
\end{align}
Similar to the $\phi^4$ theory, we compare each model’s predictions with the exact solution for the Sine-Gordon equation. 
Figure \ref{fig:SG_pred} presents a comparison between the exact solution and the predictions by NNDE and PINN for the Sine-Gordon equation. 
This plot, generated using a step size of $10^{-5}$ and a batch size of $10\%$, shows that both models closely match the exact solution. 
Similar results are seen with various selections of step size and batch size.
As shown in Table \ref{compare-table},  NNDE achieved an average MSE of 0.014344, which is close to the average MSE of 0.015208 obtained with PINN, resulting in a relative difference of only $5.68\%$. 
However, NNDE requires significantly less computation time, averaging 1702.768 seconds compared to 3144.233 seconds for PINN, representing an $45.84\%$ reduction in time.
In addition, Figure \ref{fig:SG_combined} shows the trends in accuracy and computation time for different step sizes and batch sizes, further emphasizing NNDE's advantage in computational efficiency for the Sine-Gordon equation.

\section{Summary and discussions}

In this study, we developed a novel neural network model, which we refer to as NNDE, aimed at solving differential equations. 
Specifically, we investigated the effectiveness of NNDE in finding topological soliton solutions to non-linear field equations, such as those in the $\phi^4$ theory and the Sine-Gordon equation. 
The performance of NNDE was assessed based on computational time and the accuracy of the solutions under various choices of discretization step sizes and batch sizes. 
Overall, the results show that NNDE can achieve accuracy comparable to that of PINN while significantly reducing computation time across both the $\phi^4$ theory and the Sine-Gordon equation. 

The significant difference in computation time may stem from the method used to calculate derivatives. 
Although both of NNDE and PINN use backpropagation to compute the derivative of the loss function with respect to the model parameters, 
PINN necessitates additional backpropagation steps to perform the automatic differentiation to obtain the derivatives of the target functions.
These backpropagation steps are computationally intensive, as they involve calculating gradients with respect to each parameter in the network.
Specifically, higher-order derivatives require more backpropagation steps, leading to longer computation times.
In contrast, NNDE employs a difference method to approximate derivatives, which avoids the need for repeated backpropagation steps. 
By discretizing the differential equations directly, NNDE could compute the necessary gradients more efficiently, leading to faster training times. 

While higher-order derivatives in PINN require additional backpropagation steps, NNDE necessitates the use of more neighboring points to compute the differences corresponding to these higher-order derivatives. 
This implies that NNDE consumes more GPU memory resources.

The proposed method for solving topological solitons using NN holds significant potential for the broader scientific community. 
This approach offers a novel and efficient way to address nonlinear differential equations, which play a crucial role in various fields of physics and applied mathematics. 
By reducing computational time, our method enables researchers to analyze complicated soliton dynamics using limited computational resources.
This opens up new opportunities for those working in fields such as nonlinear optics, condensed matter physics, and spintronics, where the study of solitons is critical for advancing both theoretical understanding and practical applications.




Beyond its direct applications, our work on NNDE may inspire further advancements in the study of other nonlinear systems. 
Given the universality of nonlinear dynamics in nature, from biological systems to financial models, our approach could be adapted and applied to a wide range of problems. This would expand its impact beyond physics to other fields that require the efficient analysis of complicated dynamical systems.

\section*{Acknowledgements}
The work of K. H., K. M., M. M. and G. O. was supported in part by JSPS KAKENHI Grant Nos. JP22H01217, JP22H05111, and JP22H05115

\section*{Reference}
\bibliographystyle{unsrt}
\bibliography{Reference}

\begin{thebibliography}{10}

\bibitem{MantonSutcliffe2004}
Nicholas Manton and Paul Sutcliffe.
\newblock {\em Topological Solitons}.
\newblock Cambridge University Press, 2004.

\bibitem{DrazinJohnson1989}
Philip~G. Drazin and Robin~S. Johnson.
\newblock {\em Solitons: An Introduction}.
\newblock Cambridge University Press, 1989.

\bibitem{Scott1999}
Alwyn Scott.
\newblock {\em The Nonlinear Universe: Chaos, Emergence, Life}.
\newblock Springer, 1999.

\bibitem{HORNIK1989359}
Kurt Hornik, Maxwell Stinchcombe, and Halbert White.
\newblock Multilayer feedforward networks are universal approximators.
\newblock {\em Neural Networks}, 2(5):359--366, 1989.

\bibitem{chen2018neural}
Ricky~TQ Chen, Yulia Rubanova, Jesse Bettencourt, and David~K Duvenaud.
\newblock Neural ordinary differential equations.
\newblock {\em Advances in neural information processing systems}, 31, 2018.

\bibitem{saito2018method}
Hiroki Saito.
\newblock Method to solve quantum few-body problems with artificial neural networks.
\newblock {\em Journal of the Physical Society of Japan}, 87(7):074002, 2018.

\bibitem{pfau2020ab}
David Pfau, James~S Spencer, Alexander~GDG Matthews, and W~Matthew~C Foulkes.
\newblock Ab initio solution of the many-electron schr{\"o}dinger equation with deep neural networks.
\newblock {\em Physical Review Research}, 2(3):033429, 2020.

\bibitem{hermann2020deep}
Jan Hermann, Zeno Sch{\"a}tzle, and Frank No{\'e}.
\newblock Deep-neural-network solution of the electronic schr{\"o}dinger equation.
\newblock {\em Nature Chemistry}, 12(10):891--897, 2020.

\bibitem{naito2023multi}
Tomoya Naito, Hisashi Naito, and Koji Hashimoto.
\newblock Multi-body wave function of ground and low-lying excited states using unornamented deep neural networks.
\newblock {\em Physical Review Research}, 5(3):033189, 2023.

\bibitem{hashimoto2018deep}
Koji Hashimoto, Sotaro Sugishita, Akinori Tanaka, and Akio Tomiya.
\newblock Deep learning and the ads/cft correspondence.
\newblock {\em Physical Review D}, 98(4):046019, 2018.

\bibitem{RaissiPerdikarisKarniadakis2019}
Maziar Raissi, Paris Perdikaris, and George~E. Karniadakis.
\newblock Physics-informed neural networks: A deep learning framework for solving forward and inverse problems involving nonlinear partial differential equations.
\newblock {\em Journal of Computational Physics}, 378:686--707, 2019.

\bibitem{bafghi2023pinnstf2fastuserfriendlyphysicsinformed}
Reza~Akbarian Bafghi and Maziar Raissi.
\newblock Pinns-tf2: Fast and user-friendly physics-informed neural networks in tensorflow v2, 2023.

\end{thebibliography}
\end{document}